\documentclass{ws-procs9x6-cpt19}

\def\ring#1{{\mathaccent'27 #1}}

\begin{document}

\newcommand{\refeq}[1]{(\ref{#1})}
\def\etal {{\it et al.}}

\title{Vacuum Cherenkov Radiation for Lorentz-Violating Fermions}

\author{Marco Schreck}

\address{Departamento de F\'{\i}sica, Universidade Federal do Maranh\~{a}o,\\
Campus Universit\'{a}rio do Bacanga, S\~{a}o Lu\'{\i}s - MA, 65080-805, Brazil}

\begin{abstract}
The current article reviews results on vacuum Cherenkov radiation
obtained for modified fermions.
Two classes of processes can occur
that have completely distinct characteristics.
The first one does not include a spin flip of the radiating fermion,
whereas the second one does.
A r\'{e}sum\'{e} will be given of the decay rates
for these processes and their properties.
\end{abstract}

\bodymatter

\section{Introduction}

Ordinary Cherenkov radiation in an optical medium
is emitted when a charged, massive particle
travels faster than the phase velocity of light
in that particular medium.
If the latter is the case,
the polarized atoms or molecules
in the vicinity of the particle trajectory
emit their wave trains in phase
whereupon these interfere constructively.
As a result,
coherent radiation is produced
that can be detected far away from the source.

The modified light cone or mass-shell structure
in Lorentz-violating theories
may be responsible for an energy loss of a charged, massive particle in vacuum.
As such a process shares certain characteristics
with Cherenkov radiation in macroscopic media,
it is known in the community as vacuum Cherenkov radiation.
In principle,
any vacuum endowed with a Lorentz-violating background field
in the context of the Standard-Model Extension\cite{Colladay:1998fq} (SME)
can be interpreted as a vacuum with a nontrivial refractive index.
This property explains the analogy.

Two very different scenarios exist for modified fermions\cite{Kostelecky:2013rta}
that may render vacuum Cherenkov radiation possible.
For spin-degenerate Lorentz violation,
photon emission in vacuum is possible
when the slope of the mass shell at a certain energy is larger than one.
Multiple photons can be emitted subsequently
as long as the previously mentioned condition is satisfied.
The condition fails when the particle energy
drops under a certain threshold,
whereupon the process ceases.

As long as the energy is sufficiently above the threshold,
the momentum of the emitted photon is large enough
such that the resulting recoil reverses the momentum direction
of the emitting particle.
A reversal of the momentum is directly connected to a helicity flip,
which is why photons of helicity $h=\pm 1$ can be emitted.
These photons are circularly polarized.
However,
when the energy of the radiating fermion is only slightly above the threshold,
the photons emitted are too soft to enable a helicity change.
As a consequence,
they have helicity $h=0$,
i.e.,
they are linearly polarized.

The alternative scenario of modified fermions
that allows for vacuum Cherenkov radiation
is that of spin-nondegenerate Lorentz violation.
In this case,
photon emission is possible for a certain helicity of the fermion,
i.e.,
the fermion loses energy and switches its mass shell.
The latter corresponds to a flip of spin or helicity
without a change of the momentum direction.
Therefore,
these particular processes are sometimes called helicity decays
and photons emitted in such processes are always circularly polarized.
Also,
in contrast to the previous class of processes,
helicity decays can occur for arbitrarily low energies of the radiating fermion.

In what follows,
the decay rates\cite{Schreck:2017isa} for both scenarios
will be presented and discussed.
These phenomenological results rest on a technical work\cite{Reis:2016hzu}
on the modified Dirac theory based on the nonminimal SME.

\section{Spin-degenerate Lorentz violation}

\begin{figure}[t]
\centering
\includegraphics[scale=0.30]{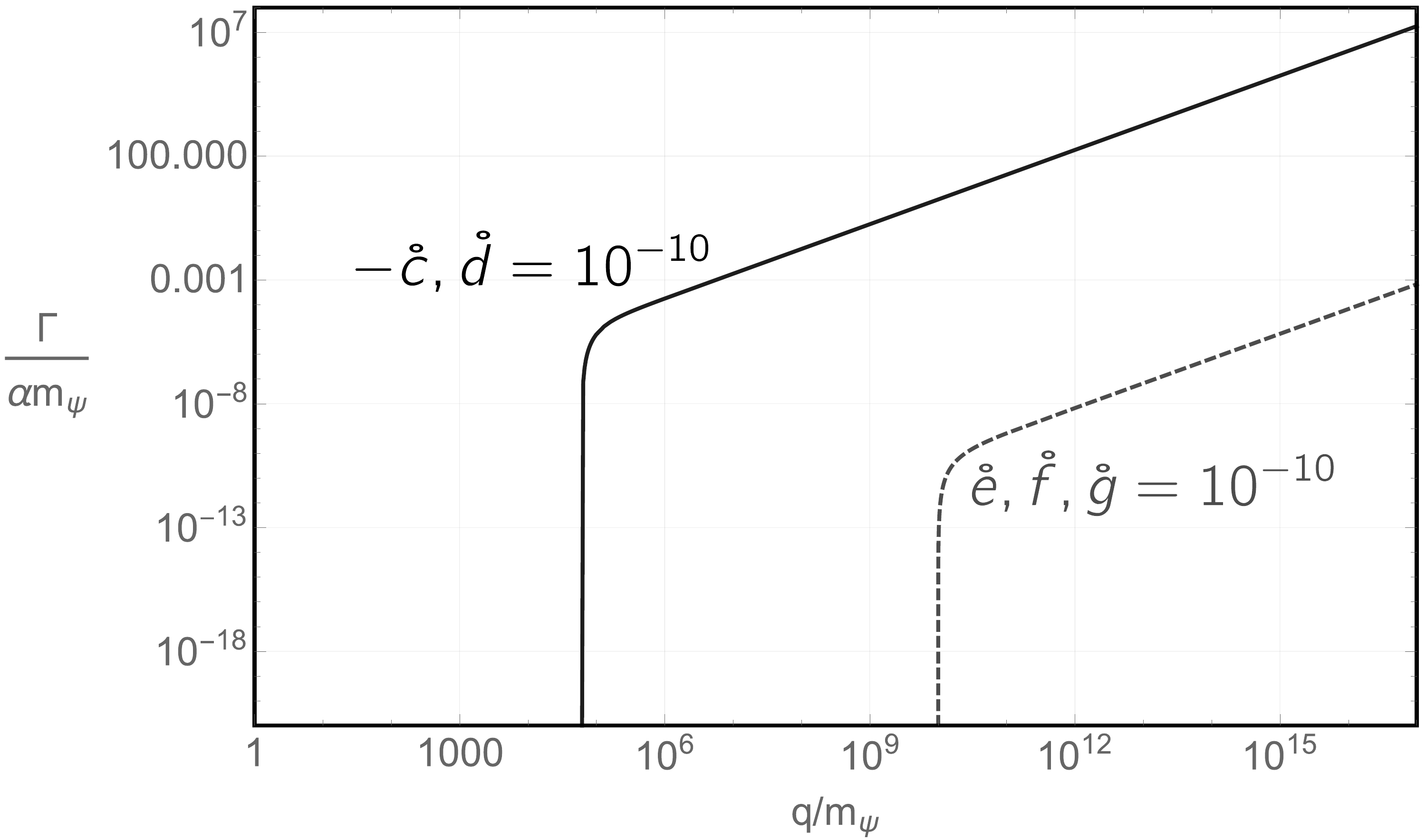}
\caption{Decay rates $\Gamma$ for the isotropic dimensionless coefficients (indicated by rings on top of the symbols) in the SME fermion sector as a function of the fermion momentum $q$ where $m_{\psi}$ is the fermion mass and $\alpha=e^2/(4\pi)$ the fine-structure constant. The plot is double-logarithmic, whereby both axes cover a large range of values.}
\label{fig:decay-rates-vacuum-cherenkov-spin-conserving}
\end{figure}

Let us first consider a fermion
modified by spin-degenerate, isotropic Lorentz violation.
The decay rates of such processes
are presented in Fig.~\ref{fig:decay-rates-vacuum-cherenkov-spin-conserving}
as functions of the incoming fermion momentum.
Minuscule values were inserted for the controlling coefficients,
as those are the most interesting ones from a phenomenological perspective.
Two curves are contained in the plot.
The first shows the behavior of the decay rates
$\Gamma_{\ring{c},\ring{d}}$
for the isotropic $c$ and $d$ coefficients,
whereas the second illustrates $\Gamma_{\ring{e},\ring{f},\ring{g}}$
for the isotropic $e$, $f$, and $g$ coefficients.
The thresholds for both curves are clearly visible
and correspond to the momenta where the rates go to zero.
For large momenta,
the decay rates grow linearly with the momentum.
In principle,
the second curve is a scaled version of the first.
The values at the axes reveal
that the $\Gamma_{\ring{e},\ring{f},\ring{g}}$ are suppressed
by one additional power of Lorentz violation
in comparison to $\Gamma_{\ring{c},\ring{d}}$.
The thresholds are correspondingly larger.
Thus,
a process of this kind can occur
for the dimensionless spin-nondegenerate coefficients $d$, $g$
whereas it is forbidden for the dimensionful coefficients $b$, $H$.

A reasonable partial crosscheck of the results can be carried out
by taking into account the coordinate transformation
that maps CPT-even, spin-degenerate Lorentz violation in the fermion sector
to CPT-even, nonbirefringent coefficients in the photon sector.\cite{Kostelecky:2009zp}
At leading order in the coefficients,
the map\cite{Altschul:2006zz} for the isotropic fermion coefficient $\ring{c}$
and the isotropic photon sector coefficient $\widetilde{\kappa}_{\mathrm{tr}}$
reads $\widetilde{\kappa}_{\mathrm{tr}}=-(4/3)\ring{c}+\dots$.
Hence,
for small enough Lorentz violation,
the decay rate\cite{Klinkhamer:2008ky} for vacuum Cherenkov radiation
in the isotropic photon sector for $\widetilde{\kappa}_{\mathrm{tr}}>0$
should correspond to the result in the isotropic fermion sector
with $\ring{c}<0$.
It was checked that the curve
that relies on the isotropic photon sector
perfectly matches $\Gamma_{\ring{c}}$.
This finding strengthens our confidence in the correctness of the results,
as both computations were carried out completely independently
from each other.
Based on the fact that ultrahigh-energy cosmic rays reach Earth,
constraints\cite{Schreck:2017isa} on Lorentz violation of quarks
embedded in a modified quantum electrodynamics were also obtained.

\section{Spin-nondegenerate Lorentz violation}

\begin{figure}[t]
\centering
\includegraphics[scale=0.30]{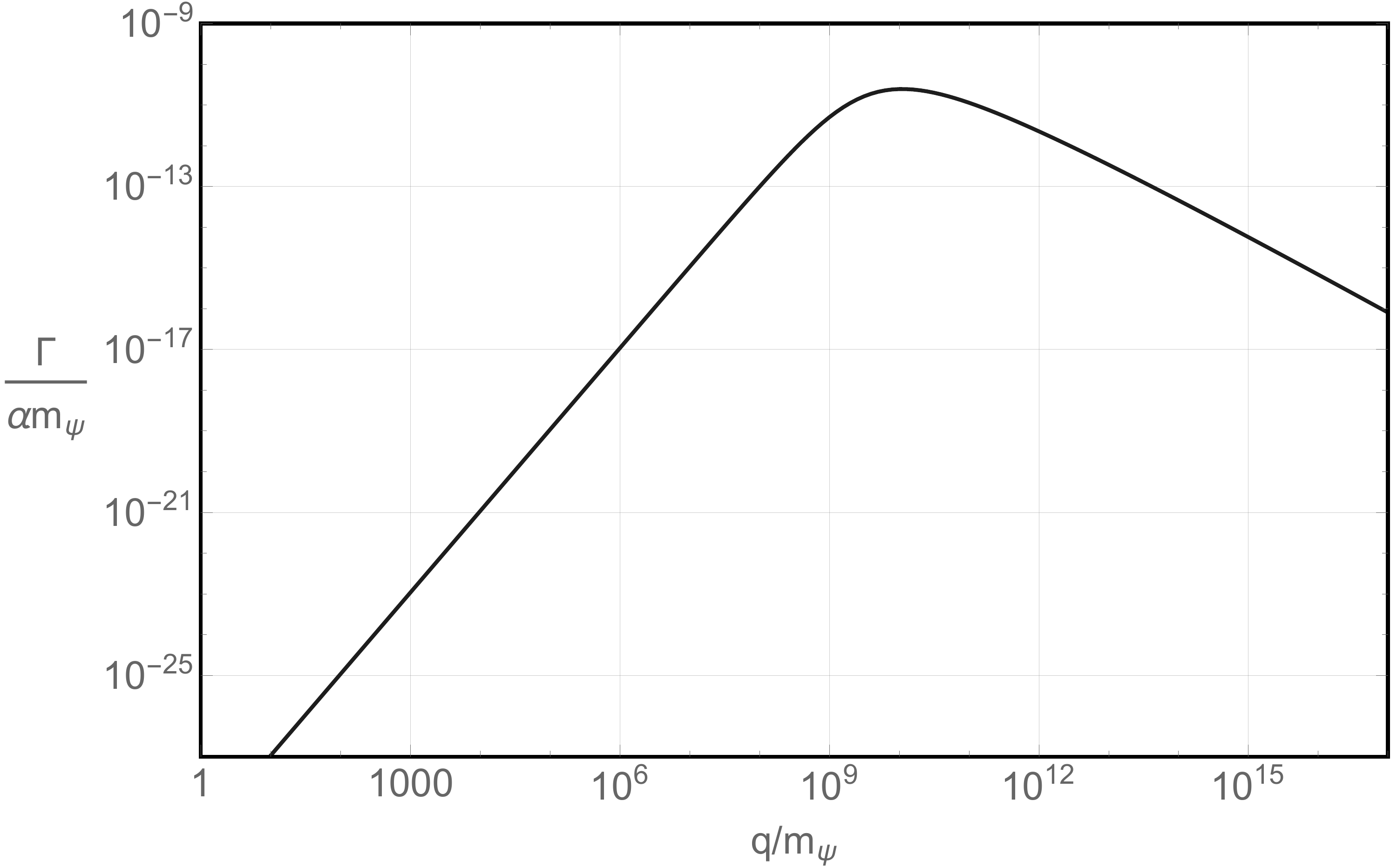}
\caption{Double-logarithmic plot of the decay rate for vacuum Cherenkov radiation for the isotropic $b$ coefficient where $\ring{b}/m_{\psi}=10^{-10}$. The quantities used are the same as those in Fig.~\ref{fig:decay-rates-vacuum-cherenkov-spin-conserving}.}
\label{fig:helicity-decay-b-isotropic}
\end{figure}

Finally,
we review the behavior of a fermion modified
by spin-nondegenerate Lorentz violation.
In particular,
the isotropic $b$ coefficient will be discussed.
The decay rate of the corresponding helicity decay
is presented in Fig.~\ref{fig:helicity-decay-b-isotropic}.
It is evident that the process does not have a threshold.
The decay rate grows polynomially
as a function of the fermion momentum
until it reaches a maximum
beyond which it decreases again logarithmically.
Furthermore,
the rate is strongly suppressed by Lorentz violation.

For high energies,
the concept of helicity becomes more and more equivalent to chirality.
Thus,
the probability of a spin flip is reduced for large energies,
which explains the decrease of the decay rate
in the high-momentum regime.
If data were available on the polarization of ultrahigh-energy cosmic rays
arriving on Earth,
fermion coefficients could,
in fact,
be constrained based on such processes.

\section*{Acknowledgments}

It is a pleasure to thank the Brazilian agencies CNPq and FAPEMA
for financial support
under grant numbers CNPq Universal 421566/2016-7,
CNPq Produtividade 312201/2018-4,
and FAPEMA Universal 01149/17.
The author greatly acknowledges the hospitality
of the Indiana University Center for Spacetime Symmetries (IUCSS).

\end{document}